\documentclass[prd,twocolumn,a4paper,floatfix,nofootinbib,preprintnumbers,superscriptaddress]{revtex4-1}

\usepackage[utf8]{inputenc}  
\usepackage[T1]{fontenc}
\usepackage{graphicx,psfrag}
\usepackage{mathrsfs}
\usepackage{amsmath,amsfonts,amssymb,pifont,gensymb}
\usepackage{multirow,enumerate}
\usepackage{comment,hyperref}
\usepackage{color}
\usepackage{acronym}
\usepackage{xspace}
\usepackage[normalem]{ulem}
\usepackage{mathtools}
\usepackage{makecell}
\usepackage{orcidlink}
\usepackage{enumitem}
\usepackage{diagbox}
\usepackage{lineno}
\usepackage{soul}

\begin{document}

\title[Revealing tensions in neutron star observations with pressure anisotropy]{Revealing tensions in neutron star observations with pressure anisotropy}

\author{Peter T. H. Pang \orcidlink{0000-0001-7041-3239}}
\email{thopang@nikhef.nl}
\affiliation{Nikhef, Science Park 105, 1098 XG Amsterdam, The Netherlands}
\affiliation{Institute for Gravitational and Subatomic Physics (GRASP), Utrecht University, Princetonplein 1, 3584 CC Utrecht, The Netherlands}

\author{Stephanie M. Brown \orcidlink{0000-0003-2111-048X}}
\affiliation{Gravitation and Astroparticle Physics Amsterdam (GRAPPA), University of Amsterdam, 1098 XH Amsterdam, The Netherlands}
\affiliation{Oskar Klein Centre, Department of Physics, Stockholm University, AlbaNova University Centre, Stockholm, SE 106 91, Sweden}

\author{Thibeau Wouters \orcidlink{0009-0006-2797-3808}}
\affiliation{Institute for Gravitational and Subatomic Physics (GRASP), Utrecht University, Princetonplein 1, 3584 CC Utrecht, The Netherlands}
\affiliation{Nikhef, Science Park 105, 1098 XG Amsterdam, The Netherlands}

\author{Chris Van Den Broeck \orcidlink{0000-0001-6800-4006}}
\affiliation{Institute for Gravitational and Subatomic Physics (GRASP), Utrecht University, Princetonplein 1, 3584 CC Utrecht, The Netherlands}
\affiliation{Nikhef, Science Park 105, 1098 XG Amsterdam, The Netherlands}

\date{\today}

\begin{abstract}
Pressure isotropy, i.e., equality between radial and tangential pressure, is often assumed when studying neutron stars. However, mechanisms such as pion/kaon condensation, magnetic fields, and dark matter clustering can lead to pressure anisotropy. This work presents a comprehensive measurement of pressure anisotropy in neutron stars. Our analysis incorporates an extensive set of nuclear experimental constraints and multi-messenger astrophysical observations. We find that the Bayes factor for anisotropy against isotropy $\gtrsim$ 3 : 1, when the anisotropy is allowed to vary between individual stars. The posterior indicates a population-wide preference for negative anisotropy, primarily driven by PSR J0740+6620. Due to the lack of radius measurements for $2M_\odot$ neutron stars, we cannot rule out density-scale-dependent anisotropy. Therefore, both phase transitions and density-scale-independent mechanisms, such as magnetic fields, dark matter clustering, or deviations from general relativity are viable explanations. While the evidence for anisotropy remains inconclusive, these results demonstrate that pressure anisotropy can be utilized as a tool for identifying missing physics in neutron star modeling or revealing novel physics in the era of multi-messenger astronomy.
\end{abstract}

\maketitle

\section{Introduction}
The multi-messenger approach to studying supranuclear matter inside neutron stars became the community standard as multi-channel observations of neutron stars became available. These observations include binary neutron star mergers, GW170817~\cite{LIGOScientific:2017vwq} and GW190425~\cite{LIGOScientific:2020aai}, detected by Advanced LIGO~\cite{LIGOScientific:2014pky} and Advanced Virgo~\cite{VIRGO:2014yos}, the electromagnetic counterpart~\cite{LIGOScientific:2017ync} associated with GW170817,  and the NICER observations of X-ray pulsars~\cite{Riley:2019yda,Miller:2019cac,Riley:2021pdl,Miller:2021qha,Vinciguerra:2023qxq,Dittmann:2024mbo,Salmi:2024aum,Mauviard:2025dmd}. The concept of ``multi-messenger'' has since been extended to include constraints from nuclear theory and experiment~\cite{Dietrich:2020efo,Huth:2021bsp,Essick:2021kjb,Legred:2021hdx,Pang:2021jta,Pang:2022rzc,Essick:2023fso,Pang:2023dqj,Tsang:2023vhh,Koehn:2024set,Legred:2025aar,Finch:2025bao}. Multi-messenger approaches have led to some of the most stringent constraints on neutron star matter to date.

In the aforementioned studies, it is commonly assumed that the pressure in the neutron star is locally isotropic, i.e., the pressure is the same in both the radial and tangential directions~\cite{Herrera:1997plx,Herrera:2020gdg}. However, there are multiple scenarios where this assumption fails, e.g., pion condensation~\cite{Sawyer:1972cq,Hartle:1974de,Sawyer:1977ja}, kaon condensation~\cite{Takatsuka:1995xn}, magnetic fields~\cite{Canuto:1968apg,Bocquet:1995je,Konno:1999zv,Weber:2006ep,Ciolfi:2010td,Gualtieri:2010md,Doneva:2012rd,Frieben:2012dz,Bucciantini:2014uia,Pili:2017yxd,Patra:2020wjy,Bordbar:2022qhl}\footnote{Other than pressure anisotropy, magnetic fields also induce additional modifications to the Tolman–Oppenheimer–Volkoff equations~\cite{Chatterjee:2018prm}, which require full numerical calculation to handle.}, superfluidity~\cite{Letelier:1980mxb,Herrera:1997plx}, dark matter clustering~\cite{Liu:2025cwy}, and violations of general relativity~\cite{Glampedakis:2015sua,daSilva:2024qex}.

Recent works have studied the existence of pressure anisotropy inside neutron stars based on observations~\cite{Biswas:2019gkw,daSilva:2024qex,Yang:2024bre,Liu:2025cwy}, but our work differs from these in several key ways. Rather than relying on a finite set of equation-of-state models, we use two flexible parameterizations and marginalize over the associated uncertainties. We incorporate a comprehensive list of nuclear experimental constraints and utilize an extensive collection of neutron stars observed through both gravitational and electromagnetic channels. Our analysis employs a robust Bayesian framework that enables statistically strong conclusions while allowing the anisotropy to vary across stars, thereby breaking the degeneracy with the equation of state. Finally, we do not assume any particular physical origin but instead demonstrate pressure anisotropy as a generic tool for probing this sector of novel physics.

\section{Methods}
\label{sec:methods}
\subsection{Barotropic equation of state}
\label{sec:EOS}

Several methods have been developed to parametrize the equation of state without assuming a specific microphysical model of supranuclear matter. These include piecewise polytropes~\cite{Read:2008iy,Steiner:2010fz,Hebeler:2013nza,Raithel:2016bux,OBoyle:2020qvf}, spectral decompositions~\cite{Lindblom:2010bb,Lindblom:2012zi,Lindblom:2018rfr,Fasano:2019zwm,Lindblom:2022mkr}, speed-of-sound parameterizations~\cite{Tews:2018iwm,Greif:2018njt}, and non-parametric approaches using Gaussian processes~\cite{Landry:2018prl,Essick:2019ldf,Landry:2020vaw} or neural networks~\cite{Han:2021kjx,Li:2025obt}. In this work, two parametric models are considered, namely, \textit{Metamodel} and \textit{Metamodel + peak}.
\subsubsection{Metamodel} 
At low densities up to $0.08 \ {\rm fm}^{-3}$, we employ the crustal equation of state from Ref.~\cite{Douchin:2001sv}. Above the crust-core transition, we assume the matter consists purely of nucleonic degrees of freedom in $\beta$-equilibrium and apply a Metamodel parametrization~\cite{Margueron:2017eqc, Margueron:2017lup, Somasundaram:2020chb}.

The Metamodel framework expresses the energy per nucleon as a systematic expansion in the isospin asymmetry,
\begin{align}
    e(n, \delta) = e_{\rm{sat}}(n) + e_{\rm{sym}}(n)\, \delta^2 + \mathcal{O}(\delta^4), 
    \label{eq:energy_per_nucleon_MM}
\end{align}
where $n$ is the baryon number density and $\delta = (n_n - n_p)/n$ is the isospin asymmetry, with $n_n$ and $n_p$ representing the neutron and proton number densities, respectively. Both the saturation energy $e_{\rm{sat}}(n)$ and symmetry energy $e_{\rm{sym}}(n)$ are expressed as Taylor expansions around saturation density $n_{\rm sat}$ using the dimensionless variable $x = (n - n_{\rm{sat}}) / 3n_{\rm{sat}}$, which in this work are truncated at second order;
\begin{align}
    e_{\rm{sat}}(n) &= E_{\rm{sat}} + \frac{1}{2} K_{\rm{sat}} x^2 + \mathcal{O}(x^3) \, ,
    \label{eq:e_sat} \\
    e_{\rm{sym}}(n) &= E_{\rm{sym}} + L_{\rm{sym}} x + \frac{1}{2} K_{\rm{sym}} x^2 + \mathcal{O}(x^3) \, .
    \label{eq:e_sym}
\end{align}
Throughout this work, we fix the saturation energy coefficient $E_{\rm{sat}} = -16 \ {\rm MeV}$ and the saturation density $n_{\rm{sat}} = 0.16 \ {\rm{fm}}^{-3}$. The remaining parameters $E_{\rm sym}$, $L_{\rm{sym}}$, $K_{\rm{sym}}$, and $K_{\rm sat}$ are referred to as the nuclear empirical parameters $\boldsymbol{\lambda}_{\rm NEP}$.

The pressure $p$ and energy density $\epsilon$ are given by,
\begin{equation}
\begin{aligned}
  p &= n^2\frac{\partial e}{\partial n}, \\
  \epsilon &= ne.
\end{aligned}
\end{equation}
For the case of symmetric matter ($\delta=0$), the saturation pressure becomes,
\begin{equation}
    p_{\rm sat}(n) = n^2\frac{\partial e_{\rm sat}}{\partial n} = \frac{n^2}{3n_{\rm sat}}K_{\rm sat}x,
\end{equation}
while the symmetry pressure is,
\begin{equation}
    p_{\rm sym}(n) = n^2\frac{\partial e_{\rm sym}}{\partial n}= \frac{n^2}{3n_{\rm sat}}\left(L_{\rm sym} + K_{\rm sym}x\right).
\end{equation}

This Metamodel framework offers an intuitive approach to incorporating laboratory constraints on nuclear matter properties, as detailed in Sec.~\ref{sec:nuclear_constraints}.

\subsubsection{Metamodel + peak}
While the Metamodel provides a flexible description of nucleonic matter, it is expected to break down when new degrees of freedom become relevant at higher densities~\cite{Lattimer:2006qiu, Oertel:2016xsn, Kumar:2023lhv}. To allow for a more flexible description of the equation-of-state and non-monotonic trend in the speed-of-sound, we extend the Metamodel using a hybrid approach similar to that described in Ref.~\cite{Greif:2018njt}.

The construction of this model has two parts. We first employ the Metamodel parametrization up to a break-off density $n_{\rm break} = 1.5 \ n_{\rm sat}$, then transition to a speed-of-sound-based description for higher densities. For densities $n > n_{\rm break}$, we parameterize the speed-of-sound squared as,
\begin{equation}
\begin{aligned}
    c^2_s &= \xi + \frac{\frac{1}{3} - \xi}{1 + e^{-l_{\rm sig}(n - n_{\rm sig})}} + c^2_{s,{\rm peak}}e^{-\frac{1}{2}\left(\frac{n - n_{\rm peak}}{\sigma_{\rm peak}}\right)^2},
\end{aligned}
\label{eq:peakCSE}
\end{equation}
where $\xi$ is determined such that the speed of sound remains continuous at $n_{\rm break}$, i.e., $c^2_s(n_{\rm break}) = c^2_{s, {\rm break}}$, with $c^2_{s, {\rm break}}$ being the speed of sound at the transition density, as determined by the Metamodel.

This parameterization incorporates two physically motivated components. The first term is a sigmoid function that captures the asymptotic approach toward the conformal limit $c^2_s=1/3$at ultra-high densities predicted by perturbative QCD calculations~\cite{Gorda:2018gpy, Gorda:2021znl, Gorda:2023mkk}. This term is characterized by a stiffness parameter $l_{\rm sig}$ and a midpoint density $n_{\rm sig}$ that control the rate and location of the transition. The second term is a Gaussian peak designed to capture intermediate-density features inferred from astrophysical observations~\cite{Legred:2021hdx, Huth:2021bsp, Pang:2021jta, Altiparmak:2022bke, Pang:2023dqj, Koehn:2024set}. This peak is defined by its maximum amplitude $c^2_{s,{\rm peak}}$, center location $n_{\rm peak}$, and width $\sigma_{\rm peak}$.

The pressure and energy density in the high-density regime are constructed by integrating the speed of sound,
\begin{equation}
\begin{aligned}
  \log\mu &= \log\mu_{\rm break} + \int^n_{n_{\rm break}} c^2_s \, d\log n',\\
  p &= p_{\rm break} + \int^n_{n_{\rm break}} c^2_s \mu \, dn',\\
  \epsilon &= \mu n - p,
\end{aligned}
\end{equation}
where $\mu$ is the chemical potential and quantities with subscript ``break'' are evaluated at $n_{\rm break}$ using the Metamodel, ensuring continuity across the transition.

Throughout this work, this hybrid construction, i.e., employing the Metamodel below $n_{\rm break}$ and the speed-of-sound parameterization above $n_{\rm break}$, is referred to as the \textit{Metamodel + peak} approach.

\subsection{Pressure anisotropy}
\label{sec:GR}
\begin{figure}[h]
    \centering
    \includegraphics[width=\columnwidth]{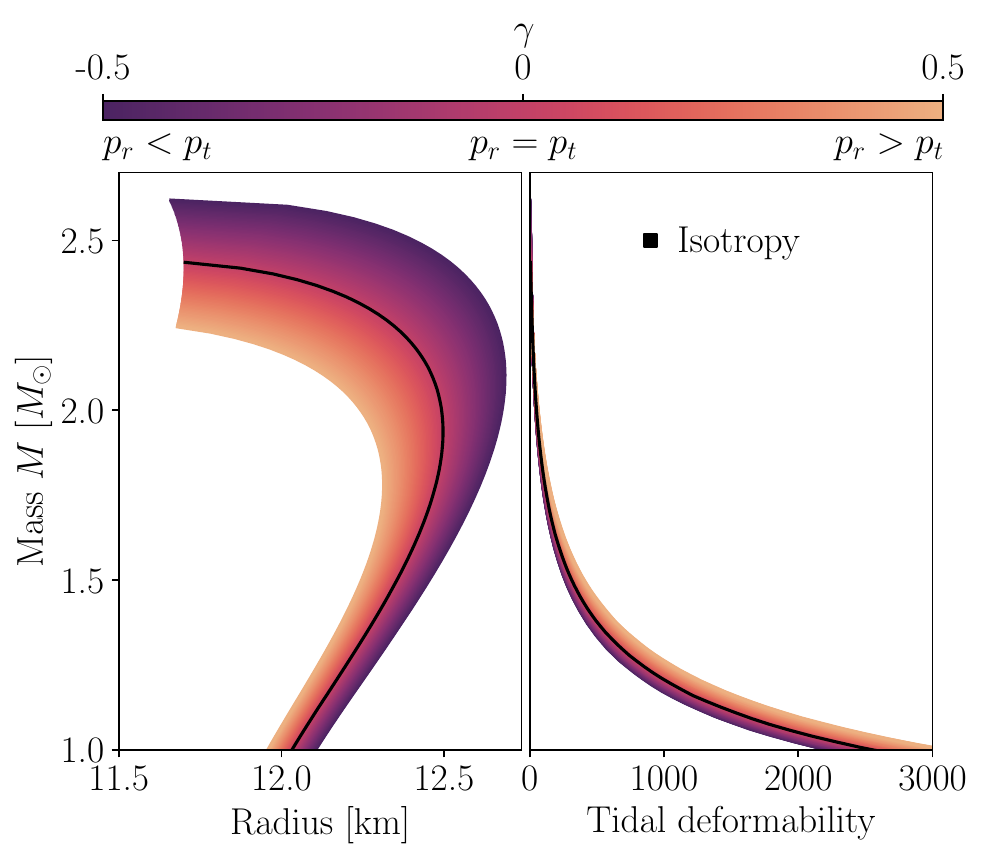}
    \caption{Illustration of the impact of pressure anisotropy on neutron star structure by varying the level of anisotropy. A positive anisotropy, i.e., the radial pressure exceeds the tangential pressure, leads to smaller yet stiffer neutron stars. In contrast, a negative anisotropy tends to produce larger but softer stars. The maximum a posteriori equation of state from Ref.~\cite{Huth:2021bsp} is used for the isotropic baseline.}
    \label{fig:eos_gamma_impact}
\end{figure}
To model the pressure anisotropy, we employ the commonly used anisotropic parameter $\sigma$, which is defined as
\begin{equation}
\label{eq:anisotropy}
    \sigma \equiv p_r - p_t,
\end{equation}
where $p_r$ is the radial pressure and $p_t$ is the tangential pressure. The inclusion of this parameter leads to a modification of the standard Tolman–Oppenheimer–Volkoff (TOV) equations. The anisotropic structure equations for static, spherically symmetric neutron stars are~\cite{Rahmansyah:2020gar},
\begin{equation}
\begin{aligned}
    \frac{dm}{dr} &= 4\pi r^2 \epsilon, \\
    \frac{dp_r}{dr} &= -\frac{m \epsilon}{r^2} \frac{\left(1 + \frac{p_r}{\epsilon}\right)\left(1 + \frac{4\pi r^3 p_r}{m}\right)}{1 - \frac{2m}{r}} - \frac{2\sigma}{r},
\end{aligned}
\label{eq:tov_mr}
\end{equation}
where $\epsilon$ is the energy density, $r$ is the radial coordinate, and $m(r)$ is the enclosed mass within radius $r$. The stellar radius $R$ is defined where the radial pressure vanishes, $p_r(R) = 0$, and the mass is $M = m(R)$.

For binary neutron stars, the component stars experience the tidal force generated by their companion. The dimensionless tidal deformability characterizes the response of a neutron star to external tidal fields,
\begin{equation}
    \Lambda = \frac{2}{3} k_2 C^{-5},
\end{equation}
where $k_2$ is the second tidal Love number and $C=M/R$ is the compactness. 

To determine $k_2$, we consider static, azimuthally symmetric metric perturbations of the form
\begin{equation}
\begin{aligned}
ds^2 = & -e^{2\Phi(r)} \left[1 + H(r) Y_{20}(\theta, \varphi) \right] dt^2 \\
& + e^{2\mu(r)} \left[1 - H(r) Y_{20}(\theta, \varphi) \right] dr^2 \\
& + r^2 \left[1 - K(r) Y_{20}(\theta, \varphi) \right] \left(d\theta^2 + \sin^2\theta\, d\varphi^2 \right).
\end{aligned}
\end{equation}
The perturbation function $H(r)$ satisfies the modified tidal equation~\cite{Biswas:2019gkw,Rahmansyah:2020gar}:
\begin{equation}
\begin{aligned}
    &H\Bigg[e^{2\mu}\left(-\frac{6}{r^2} + \frac{4\pi(p_r + \epsilon)\left(1 + \frac{d\epsilon}{dp_r}\right)}{1 - \frac{d\sigma}{dp_r}} + 4\pi(8p_r + 4\epsilon)\right)\\ 
    &+ 16\pi\sigma e^{2\mu}- (\Phi')^2\Bigg]\\
    &+ H^\prime \left[\frac{2}{r} + e^{2\mu}\left(\frac{2m}{r^2} + 4\pi r(p_r - \epsilon)\right)\right]\\
    &+ H^{\prime\prime} = 0.
\end{aligned}
\label{eq:tov_lambda}
\end{equation}

The second Love number is then given by
\begin{equation}
\begin{aligned}
    k_2 &= \frac{8 C^5}{5} (1 - 2C)^2 \left(2 + 2C(y - 1) - y\right) \\
    &\quad \times \Big[2C \left(6 - 3y + 3C(5y - 8)\right) \\
    &\qquad + 4C^3 \left(13 - 11y + C(3y - 2) + 2C^2(1 + y)\right) \\
    &\qquad + 3(1 - 2C)^2 \left(2 - y + 2C(y - 1)\right) \log(1 - 2C)\Big]^{-1},
\end{aligned}
\end{equation}
where $y= R H'(R) / H(R)$.

To solve the coupled system of Eqs.~\eqref{eq:tov_mr} and~\eqref{eq:tov_lambda}, we require closure relations connecting the radial pressure, energy density, and anisotropic parameter. For the radial pressure, we employ the barotropic equation of state models (Metamodel and Metamodel + peak) described in Sec.~\ref{sec:EOS}. The anisotropic parameter is modeled following Ref.~\cite{Doneva:2012rd}:
\begin{equation}\label{eq: definition of gamma}
    \sigma = \gamma\frac{2mp_r}{r},
\end{equation}
where $\gamma$ is a free parameter quantifying the anisotropy strength, with $\gamma=0$ corresponding to the isotropic case. This anisotropy model is preferred over the other proposed models (e.g., Ref.~\cite{Bowers:1974tgi,Herrera:2013fja}), since it satisfies the energy conditions of a perfect fluid and is symmetrically valid for both positive and negative anisotropies~\cite{Setiawan:2019ojj}, which facilitates an unbiased search for anisotropy in observations. It should be noted that such a model does not describe the actual pressure anisotropy originating from the aforementioned physical mechanisms. Rather, this model should be interpreted as a null test, i.e., to see whether pressure isotropy can be rejected when such degrees of freedom are included in the analysis.

Fig.~\ref{fig:eos_gamma_impact} illustrates the impact of non-zero pressure anisotropy on neutron star structure. Notably, a positive anisotropy, i.e., radial pressure greater than tangential pressure, leads to smaller yet stiffer neutron stars. In contrast, a negative anisotropy tends to produce larger but softer stars.

The numerical solution of these modified neutron star structural equations, combined with the equation-of-state models from Sec.~\ref{sec:EOS}, is calculated on-the-fly during the parameter estimation using the \texttt{jester} code\footnote{\url{https://github.com/nuclear-multimessenger-astronomy/jester}}~\cite{Wouters:2025zju}, which is a TOV-solver with support for GPU acceleration based on \texttt{jax}~\cite{frostig2018compiling, kidger2021on}.

\subsection{Bayesian statistics}\label{bayesian}
According to Bayes' theorem, the posterior probability $p(\boldsymbol{\theta} | d,\mathcal{H})$ on parameters $\boldsymbol{\theta}$ under the hypothesis $\mathcal{H}$ and conditioned on the data $d$ is given by
\begin{equation}
\begin{aligned}
        P(\boldsymbol{\theta} | d,\mathcal{H}) =  \frac{P(d|\boldsymbol{\theta},\mathcal{H})P(\boldsymbol{\theta}|\mathcal{H})}{P(d|\mathcal{H})} \to 
        \mathcal{P}(\boldsymbol{\theta}) =  \frac{\mathcal{L}(\boldsymbol{\theta})\pi(\boldsymbol{\theta})}{\mathcal{Z}}\,,
\end{aligned}
\end{equation}
where $\mathcal{P}(\boldsymbol{\theta})$, $\mathcal{L}(\boldsymbol{\theta})$, $\pi(\boldsymbol{\theta})$, and $\mathcal{Z}$ are the posterior, likelihood, prior, and the evidence, respectively. 
The evidence marginalizes the likelihood with respect to the prior, i.e,
\begin{equation}
    \mathcal{Z} = \int d\boldsymbol{\theta} \mathcal{L}(\boldsymbol{\theta})\pi(\boldsymbol{\theta})\,.
\end{equation}

To quantify the plausibilities between two hypotheses, $\mathcal{H}_1$ and $\mathcal{H}_2$, one can calculate the odds ratio between them, $\mathcal{O}^1_2$, given by
\begin{equation}
    \mathcal{O}^1_2 \equiv \frac{P(d|\mathcal{H}_1)}{P(d|\mathcal{H}_2)}\frac{P(\mathcal{H}_1)}{P(\mathcal{H}_2)} \equiv \mathcal{B}^1_2\Pi^1_2\,,
\end{equation}
where $\mathcal{B}^1_2$ and $\Pi^1_2$ are the Bayes factor and prior odds, respectively. 
If $\mathcal{O}^1_2 > 1$, $\mathcal{H}_1$ is more plausible than $\mathcal{H}_2$, and vice versa. Throughout this work, we set the prior odds to $1$ whenever comparing two hypotheses. Therefore, the Bayes factor is equivalent to the odds ratio.

To combine information across multiple independent observations, we adopt a hierarchical Bayesian technique. In this approach, a set of hyperparameters $\boldsymbol{\lambda}$ is introduced to describe the distribution of the parameters $\boldsymbol{\theta}$, which are specific to individual observations. We condition the prior on these hyperparameters, i.e., $\pi(\boldsymbol{\theta}|\boldsymbol{\lambda})$, and assign a hyperprior $\Pi(\boldsymbol{\lambda})$.

Given $N$ independent experiments $\{d_i\}$, the likelihood of the hyperparameters $\mathcal{L}(\boldsymbol{\lambda})$ is given by
\begin{equation}
\begin{aligned}
    \mathcal{L}(\boldsymbol{\lambda}) &= \prod_{i=1}^N \int d\boldsymbol{\theta}_i\, \mathcal{L}_i(\boldsymbol{\theta}_i)\, \pi(\boldsymbol{\theta}_i|\boldsymbol{\lambda})\\
    &=\prod_{i=1}^N \mathcal{Z}_i\int d\boldsymbol{\theta}_i\, \mathcal{P}_i(\boldsymbol{\theta}_i)\, \frac{\pi(\boldsymbol{\theta}_i|\boldsymbol{\lambda})}{\pi(\boldsymbol{\theta}_i)}\\
    &\propto\prod_{i=1}^N \int d\boldsymbol{\theta}_i\, \mathcal{P}_i(\boldsymbol{\theta}_i)\, \frac{\pi(\boldsymbol{\theta}_i|\boldsymbol{\lambda})}{\pi(\boldsymbol{\theta}_i)}
\end{aligned}
\label{eq:hyperparameter_likelihood}
\end{equation}
where $\mathcal{P}_i(\boldsymbol{\theta}_i)$ denotes the posterior corresponding to the $i$-th observation, obtained using an unconditional prior $\pi(\boldsymbol{\theta}_i)$. Throughout this work, we refer to this as the unconditioned posterior distribution.

\subsection{Nuclear experimental constraints}
\label{sec:nuclear_constraints}
To establish an informed prior on supranuclear matter, which is important for probing the anisotropy manifested in neutron stars, we first constrain nuclear parameters using terrestrial nuclear experiments. The results are then incorporated as priors on the nuclear empirical parameters when analyzing astrophysical data.

Tab.~\ref{tab:nuclear_constraints} lists the nuclear experimental constraints considered here to construct the informed prior on the nuclear empirical parameters. The constraints are similar to those in Ref.~\cite{Tsang:2023vhh}.
In the following, we present a detailed account of the constraints on the symmetric and asymmetric components of the nuclear empirical parameters.

\begin{table*}
    \caption{Nuclear experimental constraints considered. For each constraint, the uncertainty in both the density probed and the measurement is taken into account in the analysis. This table is similar to the one presented in Ref.~\cite{Tsang:2023vhh}.}
    \label{tab:nuclear_constraints}
    \centering
    \begin{tabular*}{\textwidth}{@{\extracolsep{\fill}}lccc}
        \hline
        \textbf{Symmetric matter} &&& \\
        \textbf{Constraints} & Density $n$ [fm$^{-3}$] & $K_\text{sat}$ [MeV] & $p_\text{sat}$ [MeV fm$^{-3}$]\\
        \hline
        Giant monopole resonance~\cite{Dutra:2012mb}       & 0.16 &     $230 \pm 30$ & \\
        Danielewicz et al. 2002~\cite{Danielewicz:2002pu}  & 0.32 &   & $10.1 \pm 3.0$ \\
        FOPI~\cite{LeFevre:2015paj} & 0.32 &  & $10.3 \pm 2.8$ \\
        \hline
        \textbf{Asymmetric matter} &&& \\
        \textbf{Constraints} & Density $n$ [fm$^{-3}$] & $e_\text{sym}$ [MeV] & $p_\text{sym}$ [MeV fm$^{-3}$] \\
        \hline
        Electric dipole polarizability of $^{208}{\rm Pb}$~\cite{Zhang:2015ava} & 0.05  & $15.9 \pm 1.0$  &                 \\
        PREX-II~\cite{PREX:2021umo,Reed:2021nqk,Lynch:2021xkq} & $0.11$ & & $2.38\pm0.75$ \\
        {\textbf{Nuclear masses}} &&& \\
        Brown et al. 2013~\cite{Brown:2013mga,Lynch:2021xkq} & $0.101 \pm 0.005$  & $24.7 \pm 0.8$   &         \\    
        Kortelainen et al. 2011~\cite{Kortelainen:2011ft,Lynch:2021xkq}   & $0.115 \pm 0.002$  & $25.4 \pm 1.1$   &     \\
        Danielewicz et al. 2016~\cite{Danielewicz:2016bgb,Lynch:2021xkq}  & $0.106 \pm 0.006$  & $25.5 \pm 1.1$   &     \\
        {\textbf{Heavy-ion collisions}} &&& \\
        \hspace{0.05cm}Isospin diffusion~\cite{Tsang:2008fd,Lynch:2021xkq}     & $0.035 \pm 0.011$  & $10.3 \pm 1.0$   &                \\
        \hspace{0.05cm}n/p ratio~\cite{Morfouace:2019jky,Lynch:2021xkq}   & $0.069 \pm 0.008$  & $16.8 \pm 1.2$   &                \\
        \hspace{0.05cm}n/p flow~\cite{Cozma:2017bre,Russotto:2011hq,Russotto:2016ucm,Lynch:2021xkq}    & 0.240              &                  & $12.1 \pm 8.4$ \\
        \hspace{0.05cm}$\pi$ ratio~\cite{SpiRIT:2021gtq,Lynch:2021xkq}       & $0.232 \pm 0.032$  & $52 \pm 13$      & $10.9 \pm 8.7$ \\
        \hline
    \end{tabular*}
\end{table*}

\subsubsection{Symmetric nuclear matter}
At saturation density, we adopt $K_{\rm sat} = 230 \pm 30$ MeV based on giant monopole resonance constraints~\cite{Dutra:2012mb}. Beyond saturation, collective flow measurements in high-energy Au+Au collisions provide additional constraints~\cite{Danielewicz:2002pu, LeFevre:2015paj}. Both Ref.~\cite{Danielewicz:2002pu} (using transverse and elliptical flow at $0.2-10$ GeV per nucleon) and the FOPI experiment~\cite{LeFevre:2015paj} (using elliptical flow at $0.4-1.5$ GeV per nucleon) show excellent agreement near $2 \ n_{\rm sat}$, where we take that as effective density and extract pressure constraints~\cite{Tsang:2023vhh}.

\subsubsection{Asymmetric nuclear matter}
Constraints on the symmetry energy $e_{\rm sym}(n)$ and pressure $p_{\rm sym}(n)$ come from nuclear structure studies and heavy-ion experiments across different density regimes.

Nuclear structure approaches provide density-specific constraints; at $\sim 0.1$ fm$^{-3}$ from nuclear mass data~\cite{Brown:2013mga, Kortelainen:2011ft} and isobaric analogue state energies~\cite{Danielewicz:2016bgb}, and at $0.05$ fm$^{-3}$ from $^{208}$Pb dipole polarizability measurements~\cite{Zhang:2015ava}. Moreover, the neutron skin thickness measurement by PREX-II~\cite{PREX:2021umo,Reed:2021nqk} has led to the estimation of the symmetry pressure at $0.11$ fm$^{-3}$~\cite{Lynch:2021xkq}.

Heavy-ion collisions probe broader density ranges. Low-energy collisions ($<100$ MeV per nucleon) access sub-saturation densities during post-collision expansion, with constraints from isospin diffusion in Sn+Sn collisions at $0.035\pm0.011$ fm$^{-3}$~\cite{Tsang:2008fd} and neutron-to-proton energy ratios at $0.069\pm0.008$ fm$^{-3}$~\cite{Morfouace:2019jky}. High-energy collisions ($>200$ MeV per nucleon) probe supra-saturation densities through neutron/proton elliptical flows in Au+Au collisions~\cite{Russotto:2011hq,Russotto:2016ucm,Cozma:2017bre} and charged pion ratios comparing neutron-rich and symmetric Sn+Sn systems~\cite{SpiRIT:2021gtq}.

\subsubsection{Posteriors on nuclear empirical parameters}
For these experimental constraints, the hyperparameters are the nuclear empirical parameters, $\boldsymbol{\lambda} = \boldsymbol{\lambda}_{\rm NEP}$. Each experiment has parameters $\boldsymbol{\theta}_i = \{\boldsymbol{O}_i, n_i\}$, where $\boldsymbol{O}_i$ is the observable (e.g., $p_{\rm sat}, e_{\rm sym}, p_{\rm sym}$) and $n_i$ is the density. We approximate the unconditioned posteriors as Gaussians centered at the reported values from Tab.~\ref{tab:nuclear_constraints}, with the reported uncertainties as standard deviations.

The hyperparameter conditioned priors are,
\begin{equation}
\begin{aligned}
    \pi(\boldsymbol{\theta}_i | \boldsymbol{\lambda}_{\rm NEP}) = \delta(\boldsymbol{O}_i - \boldsymbol{O}^{\rm est}_i(\boldsymbol{\lambda}_{\rm NEP}, n_i))\pi(n_i),
\end{aligned}
\end{equation}
where $\delta(x)$ is the Dirac delta, $\pi(n_i)$ is the unconditioned prior on the density probed and $\boldsymbol{O}^{\rm est}_i(\boldsymbol{\lambda}_{\rm NEP}, n_i)$ is the observable estimated at density $n_i$ based on the nuclear empirical parameters $\boldsymbol{\lambda}_{\rm NEP}$.

\begin{figure}[h]
    \centering
    \includegraphics[width=\columnwidth]{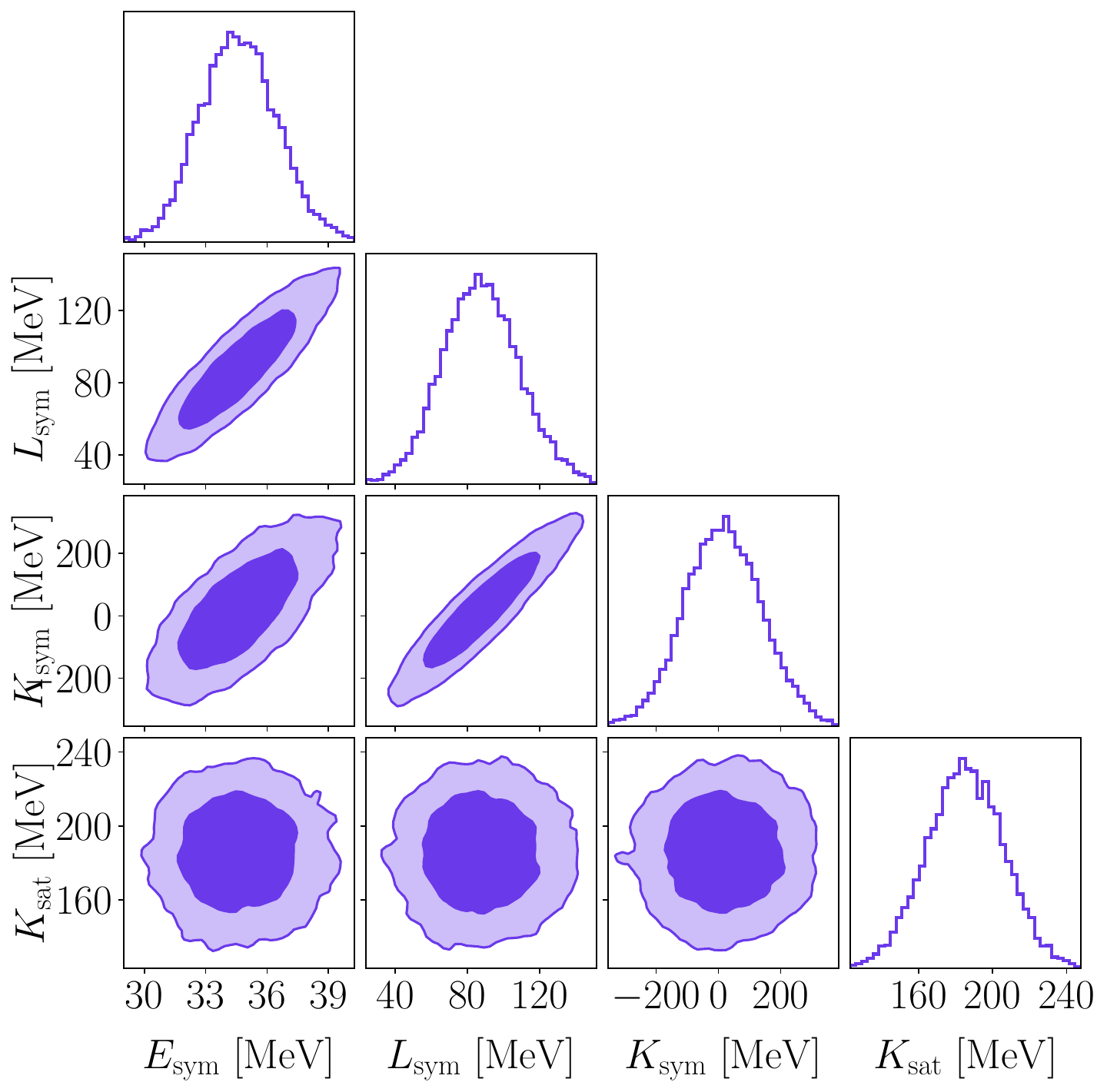}
    \caption{Posterior on the nuclear empirical parameters conditioned on the experimental constraints tabulated in Tab.~\ref{tab:nuclear_constraints}. The darker and lighter shading are indicating the $68\%$ and $95\%$ credible intervals, respectively}
    \label{fig:nep_posterior}
\end{figure}

The resulting median and $95\%$ credible intervals of the posteriors on the nuclear empirical parameters are given in Tab.~\ref{tab:nep_prior}, and the complete posterior distributions are shown in Fig.~\ref{fig:nep_posterior}. 
The resulting posterior is then approximated by a multivariate Gaussian, which allows us to account for correlations between the nuclear empirical parameters derived from the nuclear physics constraints. The resulting Gaussian distribution is then used as the prior for analyzing astrophysical observations.

\begin{table}[h]
    \caption{Prior on the nuclear empirical parameters used for analyzing nuclear experimental constraints from Tab.~\ref{tab:nuclear_constraints} and the resulting posterior. $\mathcal{U}(a, b)$ refers to a uniform distribution in $[a, b)$ and $\mathcal{N}(\mu, \sigma)$ refers to a Gaussian distribution with mean $\mu$ and standard deviation $\sigma$. The reported posterior values are the median, with the $95\%$ credible interval indicating uncertainty.}\label{tab:nep_prior}
    \begin{tabular}{@{\extracolsep{\fill}}lcc}
    \hline
    Parameter & Prior & Posterior \\
    \hline
    $E_{\rm sym}\ [{\rm MeV}]$ & $\mathcal{U}(24.7, 40.3)$ & $34.4^{+3.9}_{-3.5}$\\
    $L_{\rm sym}\ [{\rm MeV}]$ & $\mathcal{U}(-11.4, 149.4)$ & $87.3^{+43.3}_{-40.7}$\\
    $K_{\rm sym}\ [{\rm MeV}]$ & $\mathcal{U}(-400, 400)$ & $15.2^{+246.0}_{-242.4}$\\
    $K_{\rm sat}\ [{\rm MeV}]$ & $\mathcal{N}(230, 30)$ & $185.4^{+40.4}_{-40.8}$\\
    \hline
    \end{tabular}
\end{table}

\subsection{Astrophysical observations}
\label{sec:astro_constraint}
\begin{table*}
    \caption{Astrophysical observational constraints used in this work. The quoted values represent the median with $1\sigma$ credible intervals. The masses are quoted as an indication of the relevant mass scale observed. Note that for gravitational-wave events and NICER observations, the full parameter posteriors are used in the likelihood calculations rather than the summarized values presented in this table.}
     \label{tab:astro_constraints}
    \centering
    \begin{tabular*}{\textwidth}{@{\extracolsep{\fill}}lccc}
        \hline
        \textbf{Observations} & Relevant parameters & Measured values & Mass [$M_\odot$] \\
        \hline
        \textbf{Gravitational wave} &&& \\
        \hspace{0.05cm}GW170817~\cite{LIGOScientific:2017vwq,Abac:2023ujg}   & $M_1,M_2,\Lambda_1,\Lambda_2$  & \makecell[l]{$\Lambda_1=367.58^{+189.06}_{-366.28}$\\$\Lambda_2=606.34^{+279.00}_{-604.07}$}  & \makecell[c]{$M_1=1.46^{+0.05}_{-0.09}$\\$M_2=1.28^{+0.09}_{-0.04}$}\\
        \hspace{0.05cm}GW190425~\cite{LIGOScientific:2020aai,Abac:2023ujg}   & $M_1,M_2,\Lambda_1,\Lambda_2$  & \makecell[l]{$\Lambda_1=303.37^{+184.49}_{-303.23}$\\$\Lambda_2=468.72^{+252.53}_{-468.70}$}  & \makecell[c]{$M_1=1.75^{+0.05}_{-0.10}$\\$M_2=1.56^{+0.08}_{-0.06}$}\\
        \textbf{NICER} &&& \\
        \hspace{0.05cm}PSR J0030$+$0451 & $M,R$ && \\
        \hspace{0.05cm}Ref.~\cite{Vinciguerra:2023qxq}  && $R=12.77^{+1.22}_{-1.09}\,\text{km}$ & $1.34^{+0.15}_{-0.15}$\\
        \hspace{0.05cm}Ref.~\cite{Miller:2019cac}       && $R=13.00^{+1.18}_{-1.12}\,\text{km}$ & $1.44^{+0.14}_{-0.16}$\\
        \hspace{0.05cm}PSR J0740$+$6620 & $M,R$ && \\
        \hspace{0.05cm}Ref.~\cite{Salmi:2024aum}        && $R=12.49^{+0.86}_{-1.15}\,\text{km}$ & $2.07^{+0.06}_{-0.07}$\\
        \hspace{0.05cm}Ref.~\cite{Dittmann:2024mbo}     && $R=12.94^{+1.20}_{-1.60}\,\text{km}$ & $2.07^{+0.08}_{-0.09}$\\
        \textbf{Radio pulsar} &&& \\
        \hspace{0.05cm}PSR J1614$-$2230~\cite{NANOGrav:2023hde}            & $M$ & $-$ & $1.94\pm0.01$\\
        \hspace{0.05cm}PSR J0348$+$0432~\cite{Antoniadis:2013pzd,Saffer:2024tlb} & $M$ & $-$ & $1.81\pm0.04$\\
        \hline
    \end{tabular*}
\end{table*}
Our analysis incorporates the astrophysical observations of neutron stars listed in Tab.~\ref{tab:astro_constraints}. For these observations, the hyperparameters are divided into two parts: the equation of state parameters $\boldsymbol{\lambda}_{\rm EOS}$ and the anisotropy parameters $\boldsymbol{\lambda}_{\gamma}$.

The equation-of-state parameterization depends on the chosen model. For the Metamodel approach, the equation of state is fully characterized by the nuclear empirical parameters, i.e., $\boldsymbol{\lambda}_{\rm EOS} = \boldsymbol{\lambda}_{\rm NEP}$. For the Metamodel + peak approach, the parameter space expands to include the additional speed-of-sound parameters from Eq.~\eqref{eq:peakCSE}, i.e., $\boldsymbol{\lambda}_{\rm EOS} = \{\boldsymbol{\lambda}_{\rm NEP}, c^2_{s, {\rm peak}}, n_{{\rm peak}}, \sigma_{{\rm peak}}, l_{{\rm sig}}, n_{{\rm sig}}\}$.

We consider two scenarios for modeling the anisotropy parameter $\gamma$. In the \textit{Universal} scenario, all neutron stars share a common value, i.e., $\boldsymbol{\lambda}_{\gamma} = \{\gamma\}$. In the \textit{Hierarchical} scenario, each star's $\gamma$ is drawn from a truncated Gaussian population distribution with mean $\mu_\gamma$ and standard deviation $\sigma_\gamma$, i.e., $\boldsymbol{\lambda}_{\gamma} = \{\mu_\gamma, \sigma_\gamma\}$. These setups are further explained in Sec.~\ref{sec:gamma_stat_setup}.

The priors on the equation-of-state parameters and anisotropy parameters are summarized in Tab.~\ref{tab:astro_prior}.

\begin{table*}
    \caption{Prior on the equation-of-state parameters and the anisotropy parameters, where $\mathcal{U}(a, b)$ refers to a uniform distribution in $[a, b)$ and $\mathcal{N}(\mu, \sigma)$ refers to a Gaussian distribution with mean $\mu$ and standard deviation of $\sigma$.}
    \label{tab:astro_prior}
    \centering
    \renewcommand{\arraystretch}{1}
    \begin{tabular*}{\textwidth}{@{\extracolsep{\fill}}ccc}
    \hline
    Parameter & Prior & Use case\\
    \hline
    \multirow{2}{*}{\shortstack{$K_{\rm sat}$, $E_{\rm sym}$\\$L_{\rm sym}$, $K_{\rm sym}$}} 
        & \multirow{2}{*}{\shortstack{Posterior obtained\\in Sec.~\ref{sec:nuclear_constraints}}} & \multirow{2}{*}{\shortstack{Metamodel \&\\ Metamodel + peak}}\\
    & & \\
    $c^2_{s, {\rm peak}}$ & $\mathcal{U}(0.1, 1)$ & \multirow{6}{*}{Metamodel + peak}\\
    $n_{\rm peak} \ [{\rm fm}^{-3}]$ & $\mathcal{U}(0.32, 1.92)$ & \\
    $\sigma_{\rm peak} \ [{\rm fm}^{-3}]$ & $\mathcal{U}(0.016, 0.8)$ & \\
    $l_{\rm sig} \ [{\rm fm}^{3}]$ & $\mathcal{U}(0.1, 1.0)$ & \\
    $n_{\rm sig} \ [{\rm fm}^{-3}]$ & $\mathcal{U}(0.32, 5.6)$ & \\
    \hline
    $\gamma$ & $\mathcal{U}(-0.5, 0.5)$ & ${\rm Universal\ } \gamma$\\
    $\mu_\gamma$ & $\mathcal{U}(-0.5, 0.5)$ & \multirow{2}{*}{${\rm Hierarchical \ } \gamma$}\\
    $\sigma_\gamma$ & $\mathcal{U}(0, 1/\sqrt{12})$ & \\
    \hline
    \end{tabular*}
\end{table*}

\subsubsection{PSR J1614$-$2230 and PSR J0348+0432}
The radio observations on PSR J1614$-$2230~\cite{NANOGrav:2023hde} and PSR~J0348+4042~\cite{Antoniadis:2013pzd,Saffer:2024tlb} have provided a lower bound on the maximum mass of a neutron star. For these two observations, the unconditioned posterior on the pulsar mass $M$ is approximated to be a Gaussian centered at the reported values, with the reported uncertainties being the standard deviation, similar to Ref.~\cite{Dietrich:2020efo}.

The hyperparameter conditioned prior is given by
\begin{equation}
\pi(M | \boldsymbol{\lambda}) =
\begin{cases}
    \dfrac{1}{M_{\rm max}} & \text{if } M \leq M_{\rm max}, \\
    0 & \text{otherwise},
\end{cases}
\end{equation}
where the $M_{\rm max}$ is the maximum mass supported by the neutron star, given the equation of state parameter $\boldsymbol{\lambda}_{\rm EOS}$ and the anisotropic parameter $\gamma$. The unconditioned prior is assumed to be uniform.

\subsubsection{PSR J0030+0451 and PSR J0740+6620}
The Neutron Star Interior Composition Explorer (NICER) mission~\cite{2016SPIE.9905E..1HG} measures both the masses and radii of pulsars, providing crucial constraints for neutron star equation-of-state studies. NICER has provided mass and radius estimates for multiple pulsars, including two that we consider here, namely PSR J0030+0451~\cite{Miller:2019cac, Vinciguerra:2023qxq} and PSR J0740+6620~\cite{Dittmann:2024mbo, Salmi:2024aum}. To enhance measurement accuracy through improved background estimation, data from the XMM-Newton telescope~\cite{Struder:2001bh, Turner:2000jy} have been utilized for both observations.

To incorporate these observational constraints into our analysis, the unconditioned posterior on the pulsar mass $M$ and radius $R$ is approximated by a normalizing flow trained on the posterior samples released by the NICER collaboration~\cite{miller_2019_3473466, vinciguerra_2023_8239000, salmi_2024_10519473, dittmann_2024_10215109}. Details of the normalizing flow implementation can be found in Sec.~\ref{sec:nf}. The hyperparameter conditioned prior is then given by
\begin{equation}
    \pi(M, R | \boldsymbol{\lambda}) = \delta(R - R^{\rm est}(M, \boldsymbol{\lambda}_{\rm EOS}, \gamma))\pi(M),
\end{equation}
where $R^{\rm est}(M, \boldsymbol{\lambda}_{\rm EOS}, \gamma)$ represents the radius of a neutron star of mass $M$ given the equation of state parameters and anisotropy, and $\pi(M)$ is the mass prior used for obtaining the unconditioned posterior. As a systematic control, we average over multiple independent analyses. For PSR J0030+0451, we average results from~\cite{Miller:2019cac, Vinciguerra:2023qxq}, while for PSR J0740+6620, we average results from~\cite{Dittmann:2024mbo, Salmi:2024aum}. The NICER observations of PSR J0437--4715~\cite{Choudhury:2024xbk}, PSR J1231--1411~\cite{Salmi:2024bss}, and PSR J0614--3329~\cite{Mauviard:2025dmd} are excluded from this work due to insufficient systematic validation through multiple independent studies.

\subsubsection{GW170817 and GW190425}
By analyzing the gravitational-wave events GW170817~\cite{LIGOScientific:2017vwq} and GW190425~\cite{LIGOScientific:2020aai}, one can infer the masses and tidal deformabilities of the neutron stars involved in each binary system. For both GW170817~\cite{LIGOScientific:2017vwq} and GW190425~\cite{LIGOScientific:2020aai}\footnote{Note that it was also proposed that GW190425 originated from a neutron star-black hole merger~\cite{Foley:2020kus, Han:2020qmn, Kyutoku:2020xka}, but this scenario is not considered in this work.}, the unconditioned posterior on the masses $M_i$ and the tidal deformabilities $\Lambda_i$ is approximated by a normalizing flow trained on the posterior samples from~\cite{Abac:2023ujg}, which analyzed the two events using the state-of-the-art \texttt{IMRPhenomXP\_NRTidalv3} waveform model, and showed consistent results with previous analyses.

The hyperparameter conditioned prior is given by
\begin{equation}
\begin{aligned}
    \pi(M_1, M_2, \Lambda_1, \Lambda_2 \mid \boldsymbol{\lambda}) &= \delta(\Lambda_1 - \Lambda_1^{\rm est}(M_1, \boldsymbol{\lambda}_{\rm EOS}, \gamma))\\
    &\times\delta(\Lambda_2 - \Lambda_2^{\rm est}(M_2, \boldsymbol{\lambda}_{\rm EOS}, \gamma))\\
    &\times\pi(M_1,M_2),
\end{aligned}
\end{equation}
where $\Lambda^{\rm est}(M, \boldsymbol{\lambda}_{\rm EOS}, \gamma)$ is the tidal deformability of a neutron star of mass $M$ given the equation of state and the anisotropy. $\pi(M_1, M_2)$ is the prior on the masses used for obtaining the unconditioned posterior.

As shown above, the anisotropy of the two neutron stars in a binary is assumed to be the same. However, this assumption may not hold if we consider variations between different stars. For the current generation of detectors, only the leading mass-weighted tidal deformability is measurable, while the individual deformabilities of the stars are not~\cite{Wade:2014vqa}. Therefore, we conclude that using a common anisotropic parameter within a binary system is sufficient.

We exclude GW190814~\cite{LIGOScientific:2020zkf} from our analysis, as there is no conclusive evidence for a matter contribution, i.e., the secondary object may be a neutron star or a black hole. Similarly, GW230529~\cite{LIGOScientific:2024elc} is omitted due to ambiguity in the nature of the primary object and the absence of tidal signatures. Finally, GW200105 and GW200115~\cite{LIGOScientific:2021qlt} are excluded, as no meaningful constraints on the tidal deformability of the secondary objects are available in either case.

\subsection{Statistical scenarios for anisotropy}
\label{sec:gamma_stat_setup}
We implement two scenarios to parameterize the anisotropy in the neutron star population. In the \textit{Universal $\gamma$} scenario, all neutron stars share a universal anisotropy parameter $\gamma$, representing a single global parameter characterizing anisotropic strength. The \textit{Hierarchical $\gamma$} scenario allows variation between neutron stars by drawing each star's anisotropy parameter $\gamma_i$ from a population-level hyper-distribution. We model this as a truncated Gaussian with mean $\mu_{\gamma}$ and standard deviation $\sigma_{\gamma}$. This scenario effectively removes the assumption that all neutron stars follow a single mass-radius-tidal deformability relation, while keeping the assumption that all neutron stars are made of the same matter. For both scenarios, the universal $\gamma$ or individual $\gamma_i$ are bounded to $[-0.5, 0.5]$ for numerical stability while permitting both positive and negative anisotropies. At $\vert\gamma\vert > 0.5$, numerical instabilities introduce non-trivial features in the prior distribution, which compromise the interpretation of our results as an unbiased null test of pressure isotropy. This bound therefore ensures that we can search for anisotropies of either sign without introducing numerical artifacts.

\section{Results}
Bayes factors comparing models with anisotropy to those without, under each equation-of-state model, are presented in Tab.~\ref{tab:bayes_factor}. In the ${\rm Universal\ } \gamma$ scenario, the results show no preference for or against the presence of anisotropy when using either of the equation-of-state models. In contrast, under the more flexible ${\rm Hierarchical \ } \gamma$ scenario, both equation-of-state models exhibit a modest preference for the inclusion of anisotropy. This effect is more pronounced for the Metamodel + peak model, suggesting that models that allow for variations in anisotropy between individual stars are more supported by the data.

\begin{figure}[h]
    \centering
    \includegraphics[width=\columnwidth]{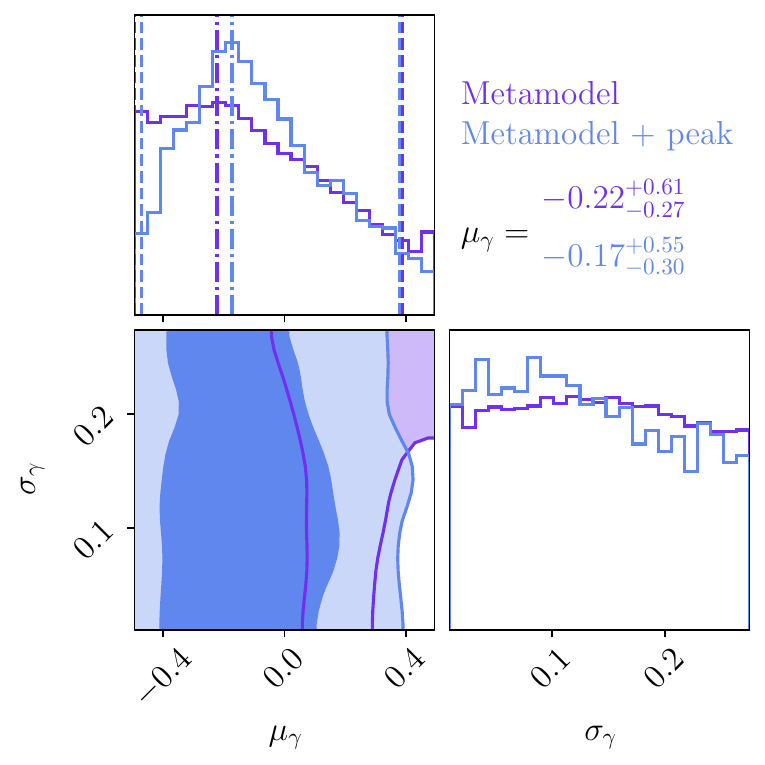}
    \caption{Posterior of $\mu_{\gamma}$ and $\sigma_{\gamma}$ for the two equation-of-state models considered. The posteriors are highly consistent and indicate a general negative anisotropy. The values quoted for $\mu_\gamma$ are the maximum a posteriori (dot-dashed line), together with the $95\%$ credible interval (dashed line)as the uncertainty. The 2D contours are shown at $68\%$ (darker shade) and $95\%$ (lighter shade) levels.}
    \label{fig:mu_gamma_posterior}
\end{figure}

Although the evidence for pressure anisotropy in neutron stars remains inconclusive, the posterior distributions reveal a systematic trend across the population. The posteriors for the hyperparameters $\mu_{\gamma}$ and $\sigma_\gamma$ are shown in Fig.~\ref{fig:mu_gamma_posterior}. While the distribution for $\sigma_\gamma$ remains uninformative, there is a clear preference for negative values of $\mu_\gamma$, indicating that a negative anisotropy is favored across the neutron star population.

Moreover, the posterior distributions obtained from both equation-of-state models show remarkable consistency. This agreement leads to two important conclusions. First, pressure anisotropy does not exhibit strong degeneracy with the equation of state when analyzing a population of neutron stars. Second, the observed trend of anisotropy is robust against the complexity of the equation-of-state model. This means it cannot be attributed to the limitations of simpler equation-of-state descriptions, as even more sophisticated models yield consistent results.

\begin{table}
  \caption{Bayes factors on the presence of pressure anisotropy against the absence of it with the two equation-of-state models considered.}
  \label{tab:bayes_factor}
  \begin{tabular}{lcc}
    \hline
    &${\rm Universal\ } \gamma$&${\rm Hierarchical\ } \gamma$\\
    \hline
    Metamodel & $1.00\pm0.04$ & $1.40\pm0.02$\\
    Metamodel + peak & $1.00\pm0.06$ & $2.73\pm0.16$\\
    \hline
  \end{tabular}
\end{table}

\begin{figure}[h]
    \centering
    \includegraphics[width=\columnwidth]{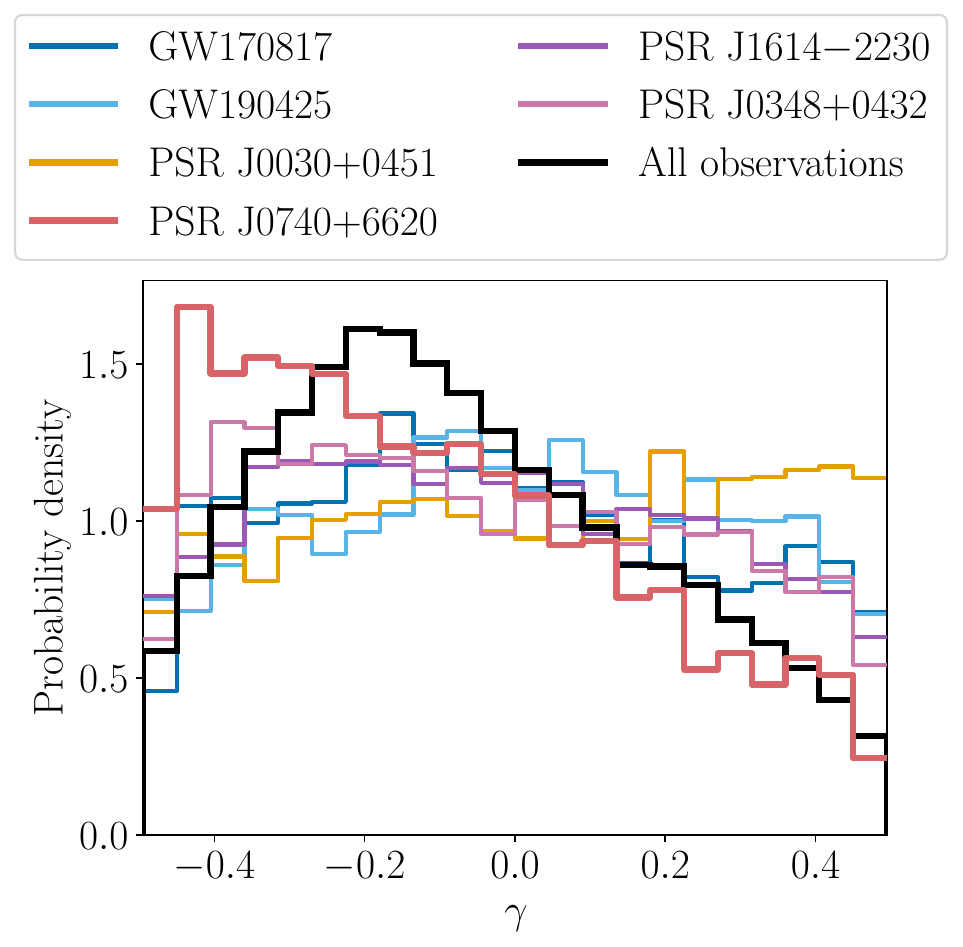}
    \caption{The posteriors of $\gamma$ for each individual observation, using the Metamodel + peak equation-of-state model are shown. The baseline, shown in black, is the predictive posterior on $\gamma$ based on the posterior of $\mu_\gamma$ and $\sigma_\gamma$ as shown in Fig.~\ref{fig:mu_gamma_posterior}, which uses all the constraints. The posteriors of individual events are relatively uninformative across all observations, except for PSR J0740+6620 (shown in red), which has a stronger preference for negative anisotropy compared to others.}
    \label{fig:individual_gamma}
\end{figure}

To better understand the source of negative anisotropy in the population, we first repeat the analysis excluding all nuclear experimental constraints. The resulting posterior remains largely consistent with that shown in Fig.~\ref{fig:mu_gamma_posterior}, yielding $\mu_\gamma=-0.27^{+0.65}_{-0.22}(-0.17^{+0.46}_{-0.31})$ for the Metamodel (Metamodel + peak). This consistency indicates that the observed anisotropy is not driven by tension between nuclear experiments and astrophysical observations.

We then examine the posterior distribution of $\gamma$ for each individual astrophysical observation. It should be noted that when examining individual observations, the conclusion will be equation-of-state model dependent, as choosing a particular model and its associated prior implicitly takes observations more aligned with it as the reference. Using the posterior distributions of $\mu_\gamma$ and $\sigma_\gamma$, we derive the resulting distribution of $\gamma$ shown in Fig.~\ref{fig:individual_gamma}. This distribution represents the baseline constraint on $\gamma$ from the combined dataset of all observations and experimental constraints, providing a reference for evaluating the individual contributions from each observation. Then we repeat the analysis, treating $\gamma$ for each observation as an independent parameter as in the Hierarchical $\gamma$ scenario, but without imposing the Gaussian hyper-distribution on them. As shown in Fig.~\ref{fig:individual_gamma}, the posterior distributions of $\gamma$ for most neutron stars are largely uninformative and consistent with a uniform distribution. Quantitatively, the Jensen–Shannon divergence between a uniform distribution and individual observations is less than $5\times10^{-3} \ {\rm bit}$ for all cases except PSR J0740+6620, which exhibits a stronger preference for negative anisotropy with a Jensen–Shannon divergence of $3\times10^{-2} \ {\rm bit}$. This result indicates that the observed radius of J0740+6620 is larger than predicted by nuclear experimental constraints and other neutron star observations, thus favoring a correction toward negative anisotropy.

\section{Discussion}
\label{sec:discussion}
Fig.~\ref{fig:individual_gamma}, along with the Bayes factors in Tab.~\ref{tab:bayes_factor}, shows that allowing anisotropy to vary between neutron stars slightly favors pressure anisotropy against isotropy, regardless of the equation of state, with the population exhibiting overall negative anisotropies. By examining the posterior of $\gamma$ for individual observations, we conclude that PSR J0740+6620 is the primary driver of this negative trend. Since negative anisotropy means that the radial pressure exceeds the tangential pressure, PSR J0740+6620 shows a weaker radial pressure than expected from nuclear constraints and other neutron star observations.

Since the detailed calculations and follow-up work necessary to determine the cause of this apparent anisotropy are beyond the scope of this work, we instead aim to provide qualitative estimates of whether such phenomena can be explained by various physical mechanisms.

When the anisotropy is allowed to vary between stars, there is stronger statistical evidence for anisotropy than when there is a single, universal value. This statistical preference, however, is mild. Additionally, due to the lack of radius measurements for neutron stars around $2M_{\odot}$\footnote{We note that there are also observations for the X-ray burster 4U 1702-429~\cite{Nattila:2017wtj}, which provide radius measurements around the $2M_\odot$ mass scale, but the systematic uncertainties remain non-negligible and are not considered in this work.}, it remains unclear whether there is a correlation with mass and, correspondingly, with density scales. As a result, potential explanations,  such as a phase transition or density-induced superfluidity, cannot be ruled out. Pion condensation, in particular, is characterized by a mass-dependent negative anisotropy. In contrast, the impacts of strong magnetic fields, dark matter clustering, and violations of general relativity are not expected to correlate strongly with the mass scale. Strong magnetic fields lead to negative anisotropy and more massive stars, while both dark matter and violations of general relativity can cause positive or negative anisotropy depending on the exact mechanism. All of these mechanisms, therefore, remain viable explanations.

We emphasize that the above interpretations are speculative and assume the presence of pressure anisotropy, which remains only weakly supported by current data. However, these findings indicate that anisotropy serves as a valuable generic tool for probing missing sectors in multi-messenger neutron star studies. Because anisotropy is not strongly degenerate with equation-of-state parameters, it allows us to probe individual variations and identify potential tensions between observations.

\section{Conclusion}
\label{sec:conclusion}
In this work, we conducted a comprehensive measurement of pressure anisotropy by utilizing an extensive set of nuclear experimental constraints and astrophysical observations of neutron stars. By employing a robust Bayesian framework, we are able to perform this measurement while effectively marginalizing over the uncertainties of the supranuclear equation of state. Additionally, we arrived at a statistically robust conclusion regarding the existence of this anisotropy.

Our findings indicate that across the various equation of state models considered, there is slight evidence of anisotropy when it is allowed to vary between neutron stars. The posterior analysis suggests an overall negative anisotropy across the population driven by PSR J0740+6620 rather than by a tension between the nuclear experiments and the astrophysical observations. Due to the lack of radius measurements for neutron stars around $2M_\odot$, we cannot determine whether this individual variability is due to density-scale-dependent anisotropy. Therefore, it could be attributed to a phase transition or density-independent mechanisms, such as magnetic fields or violations of general relativity.

However, given the weak statistical evidence, we cannot definitively conclude that anisotropy has been observed. Nevertheless, due to the variety of physical origins for anisotropy and its weak degeneracy with the equation of state, we propose using anisotropy as a tool for probing missing physics in modeling neutron star observations. This approach could help us investigate individual variations and identify potential novel physics among observations.

\appendix
\section{Numerical methods and implementation}
\subsection{Normalizing flow}
\label{sec:nf}
A key step in evaluating the likelihood of the hyperparameters is to estimate the unconditioned posterior distribution $\mathcal{P}(\boldsymbol{\theta})$. The posterior distribution is estimated from a given set of posterior samples using \textit{normalizing flow}~\cite{Kobyzev:2019ydm, Papamakarios:2019fms}.

Normalizing flows are a neural network-based method designed to learn a bijection transformation $f:\mathbb{R}^d\to \mathbb{R}^d$ from a latent distribution $\boldsymbol{z} \sim p_Z$, which is typically simple and easy to sample, to the true distribution. The true complicated and unknown distribution $\boldsymbol{x} \sim p_X$ is represented by samples $\{x_i\}$. After training, the model allows efficient sampling from and computation of the true distribution's probability density, i.e.,
\begin{equation}
\begin{aligned}
    \boldsymbol{x} &= f(\boldsymbol{z}),\\
    p_X(\boldsymbol{x}) &= p_Z(f^{-1}(\boldsymbol{x})) \left \vert \frac{\partial f^{-1}(\boldsymbol{x})}{\partial \boldsymbol{x}} \right\vert.
\end{aligned}
\end{equation}

In order to estimate and sample from the unconditioned posterior distribution, we employ a masked auto-regressive flow~\cite{Kingma:2016wtg, Papamakarios:2017tec}, as implemented in \texttt{FlowJAX}~\cite{ward2023flowjax}.

\subsection{Markov chain Monte Carlo}
To obtain posterior samples of the hyperparameters, we use \textit{Markov Chain Monte Carlo} (MCMC) sampling, specifically the Metropolis-Hastings algorithm. In MCMC parameter estimation, the posterior is approximated by a collection of walkers. The walkers will converge asymptotically to the posterior distribution $\mathcal{P}(\lambda)$. 

At each iteration, the sampler proposes a new point $\lambda^\prime$ based on the current state $\lambda_t$ and a proposal density $Q(\lambda' | \lambda_t)$. The proposed point is accepted with a probability $a$,
\begin{equation}
    a = \min\left(1, \frac{\mathcal{P}(\lambda')}{\mathcal{P}(\lambda_t)}\frac{Q(\lambda_t | \lambda')}{Q(\lambda' | \lambda_t)}\right).
\end{equation}
If the proposal is rejected, the chain remains unchanged.

This process is repeated until the chain has sufficiently explored the posterior distribution. The resulting samples are then treated as representative draws from the posterior distribution. This analysis utilizes the machine learning-enhanced MCMC algorithm implemented in \texttt{flowMC}~\cite{Wong:2022xvh, Gabrie:2021tlu}, in which the proposal distribution is trained on the fly using preliminary posterior samples.

\subsection{Learned harmonic mean estimator}

The \textit{learned harmonic mean estimator}~\cite{Polanska:2024arc, Piras:2024dml, SpurioMancini:2022vcy} is used to estimate the evidence given a set of posterior samples and their corresponding likelihood and prior values. This method offers a scalable, sample-based approach to evidence estimation that is independent of the methodology used to obtain the posterior samples. While the original harmonic mean estimator is known to suffer from instabilities~\cite{Newton:1994wlb}, the learned version mitigates these issues using machine learning techniques~\cite{Polanska:2024arc}.

In particular, the reciprocal of the evidence, \( \mathcal{Z}^{-1} \), is estimated via:
\begin{equation}
    \mathcal{Z}^{-1} = \widehat{\mathbb{E}}_{\boldsymbol{\theta} \sim \mathcal{P}(\boldsymbol{\theta})}\left[ \frac{\varphi(\boldsymbol{\theta})}{\mathcal{L}(\boldsymbol{\theta})\pi(\boldsymbol{\theta})}\right],
\end{equation}
where $\widehat{\mathbb{E}}_{x \sim p(x)}[\cdot]$ denotes the sample average under the distribution $ p(x)$. $\varphi(\boldsymbol{\theta})$ represents a learned, normalized target distribution that is required to concentrate within the true posterior. As demonstrated in Ref.~\cite{Polanska:2024arc}, normalizing flows offer a robust approach to ensure this concentration condition is met.

We use the machine learning-assisted harmonic mean estimator implemented in \texttt{harmonic}~\cite{astroinformatics:harmonic}.

\section*{Data availability}
The datasets generated and/or analysed during the current study are available from the corresponding author and on Ref.~\cite{pang_2026_18978514}. The NICER posterior samples used are obtained from Zenodo (\url{https://doi.org/10.5281/zenodo.3473466}, \url{https://doi.org/10.5281/zenodo.8239000}, \url{https://doi.org/10.5281/zenodo.10519473}, and \url{https://doi.org/10.5281/zenodo.10215109}).

\section*{Code availability}
All computational methods used in this work are based on open-source and publicly available software. The relativistic TOV solver is implemented in \texttt{jester} (\url{https://github.com/nuclear-multimessenger-astronomy/jester})~\cite{Wouters:2025zju}. Normalizing flows used to estimate the unconditioned posterior distribution are implemented using \texttt{FlowJAX} (\url{https://github.com/danielward27/flowjax})~\cite{ward2023flowjax}. Posterior sampling is performed using the machine learning-enhanced MCMC sampler \texttt{flowMC} (\url{https://github.com/kazewong/flowMC})~\cite{Wong:2022xvh, Gabrie:2021tlu}. Bayesian evidence is estimated using the learned harmonic mean estimator implemented in \texttt{harmonic} (\url{https://github.com/astro-informatics/harmonic})~\cite{astroinformatics:harmonic}.\\

\section*{Acknowledgment}
We thank Isaac Legred, Micaela Oertel, and the LIGO-Virgo KAGRA extreme matter group for their fruitful discussions and feedback. P.T.H.P. is supported by the research program of the Netherlands Organization for Scientific Research (NWO) under grant number VI.Veni.232.021. S.M.B. is supported by the research project grant `Fundamental physics from populations of compact object mergers' funded by VR under Dnr 2021-04195, the research project grant `Gravity Meets Light' funded by the Knut and Alice Wallenberg Foundation under Dnr KAW 2019.0112, and by the Netherlands Organization for Scientific Research (NWO) under grant number VI.Veni.242.361. T.W. and C.V.D.B. are supported by the research program of the Netherlands Organization for Scientific Research (NWO) under grant number OCENW.XL21.XL21.038. This material is based upon work supported by the NSF's LIGO Laboratory, a major facility fully funded by the National Science Foundation.

\bibliography{main}

\end{document}